\documentclass[12pt]{article}

\author{Vladimir A. Petrov \footnote{e-mail: Vladimir.Petrov@ihep.ru}  }

\title{ Coulomb-Nuclear Interference:                                                   the Latest Modification\footnote{To be published in Vol.309 of the Proceedings of the Steklov Institute of Mathematics. (Collected papers. On the occasion of the 80th birth of Academician Andrei Alekseevich Slavnov.)}  }
\date{}

\begin{document}

\maketitle
\begin{center}
A. A. Logunov Institute for High Energy Physics 

NRC "Kurchatov Institute", Protvino, RF
\end{center}
\begin{abstract}
We argue that the account of Coulomb-nuclear interference (CNI) in the differential cross-section of elastic $ pp $ scattering may be easily treated without introduction of intermediate IR regularization ("photon mass"). We also indicate that the parametrization used earlier misses some terms of the second order in $ \alpha $  while it contains a superfluous term of the first order.
\end{abstract}

\section{Introduction}	

The basis of the modern theory of strong interactions is Quantum Chromodynamics, a gauge quantum theory of quark and gluon fields which Professor A. A. Slavnov has made fundamental contributions \cite{Sla}  to. 

Our present view of high energy scattering of hadrons is dominated by the idea of a leading Regge trajectory, the Pomeron, which embodies 
colourless gluon exchanges and leads asymptotically to the hypothesis (ascendant to the celebrated Pomeranchuk theorem and later pushed forward by V. N. Gribov in early 1970s) of universal C-even ("C" means "crossing") behaviour of cross-sections independent of flavours of colliding hadrons. At low energy this is violated by "usual" quarkic  reggeons which, however, die off with energy. 

Afterwards, it was argued that besides quarkic C-odd reggeons one can admit a C-odd partner of the Pomeron, "the Odderon", which can , potentially, violate the above said universality even at high energies \cite{Luk} .

Recent measurements by the LHC TOTEM Collaboration at 13 TeV \cite{TOT} caused a vivid discussion (more than 60 publications by now) of a strikingly small value of the parameter $ \rho = {Re T_{N}(s,0)}/{Im T_{N}(s,0)}$ (here $T_{N}(s,t)$ stands for the $ pp $ scattering amplitude)   which lies (with some variations) near $ 0.10 $. It was considered in Ref.\,\cite{Nic}     as manifestation of so-called "maximal Odderon" which is to violate the strong interaction universality in a maximal possible way. 

The extraction of this $ \rho $-parameter (which, let us recall, is inherently model dependent) from the data depends decisively on how the Coulomb contributions are taken into account in the full scattering amplitude.
\section{From Bethe to CKL}
 During quite a long time the Bethe formula \cite{Be}  for the total amplitude $ T_{C+N} $ has been widely applied for extraction of the parameter $ \rho $ from the data (which is defined by $\mid T_{C+N}\mid^{2}$, see Eq.(2)) : 
 \begin{equation}
 T_{C+N} = \frac{8\pi s \alpha \mathcal{F}^{2}(t)}{t} + e^{i\alpha\Phi (s,t)} T_{N}(t)
 \end{equation}
 where $ \mathcal{F} $ is the proton e.m. form factor and $ \Phi (s,t) $ is the Bethe phase usually in the form given to it by West and Yenni \cite{We} (or some later modifications of it).
 
However, over recent years the general practice in the TOTEM publications on this subject is based, instead of Eq.(1), on the use of the Cahn-Kundr\'{a}t-Lokaj\'{i}\v{c}ek (CKL) formula \cite{Cahn}  for account of CNI  which is more general (e.g., it does not imply the $ t $  independence \footnote{Problems with $ t $ dependence of the nuclear phase were analyzed in \cite{Pet}. } of the nuclear phase $ Arg T_{N} (s,t) $ ) than the Bethe formula. 

The CKL approximation used in \cite{TOT} (in a bit different normalization) has the form \footnote{The damping factors due to the soft and virtual photons are well known but negligible in the region of CNI.}
\begin{equation}
\frac{d\sigma_{C+N}}{dt}= \frac{(\hbar c)^{2}}{16\pi s^{2}} \mid T_{C+N}\mid^{2}= \frac{(\hbar c)^{2}}{16\pi s^{2}} \mid \frac{8\pi\alpha s}{t}\mathcal{F}^{2}(t) + T_{N} [1-i\alpha G(t)]\mid^{2}
\end{equation}
with
\begin{equation}
G(t)= \int dt^{'} log (\frac{t^{'}}{t}) \frac{d}{dt^{'}} \mathcal{F}^{2}(t^{'})-\int dt^{'}(\frac{T_{N}(t^{'})}{T_{N}(t)} - 1) \frac{I(t,t^{'})}{2\pi}
\end{equation}
where $ \mathcal{F}(t) $ is the proton electric form factor and 
\begin{center}
$ I(t,t^{'})= \int_{0}^{2\pi}d\phi\: \mathcal{F}^{2}(t^{''})/t^{''},\: t^{''}=t+t^{'} + 2\sqrt{t t^{'}} \cos\phi . $
\end{center}

It is clear, however, that for proper accounting of powers of $ \alpha $ in perturbative QED expansion used in Eq.(2) one has to retain not only order $ \alpha^{1} $ terms but also terms $\sim \alpha^{2} $ . Otherwise, we will miss some terms  $\sim \alpha^{2} $ in the differential section.

Eqs. $(2)\:-\:(3)$ were obtained as a result of rather questionable manipulations \cite{Cahn} with  the IR regulator mass prior it could be finally eliminated. Eq.(2) was criticized in Ref.   \cite{Petr} where it was argued, in particular, that the term $ \int dt^{'} log (\frac{t^{'}}{t}) \frac{d}{dt^{'}} \mathcal{F}^{2}(t^{'}) $ is superfluous.
\section{Modified form of the CNI account}

To proceed further we have to notice that many problems can be overcome much easier if we realize that the square of the amplitude is \textit{free from Coulombic IR divergences}.

 Below we will use, instead of $ t $, a more convenient variable 
\begin{center}
$  q^{2} \equiv q^{2}_{\perp} = ut/4k^{2} = k^{2}sin^{2}\theta, \; s= 4k^{2}+ 4m^{2},  $
\end{center}
which reflects the $ t-u $ symmetry of the $ pp $ scattering.
At $ \theta\rightarrow 0\;\quad q^{2} \approx -t $ while at $ \theta\rightarrow \pi\;\;\quad q^{2} \approx -u $.
We will use the same notation $ q $ both for 2-dimensional vectors $ \textbf{q} $ and their modules $ \vert \textbf{q} \vert\ $. In the latter case, the limits of integration are indicated explicitly. As we deal with high energies and have in the integrands fast decreasing nuclear amplitudes and form factors we can (modulo vanishingly small corrections) extend the integration
in $ \textbf{q} $ (kinematically limited by $ \mid \textbf{q} \mid \leq \sqrt{s}/2 $ ) over the whole 2D space. The benefit is the possibility to freely use direct and reversed 2D Fourier transforms.

Thus, based on the same premises as CKL ( the additivity of the eikonal w.r.t. strong and electromagnetic interactions)  we  have obtained the following expression for the \textit{modulus squared}  of the full amplitude (i.e. for the \textit{ observed} quantity) which from the very beginning is free from IR regulators (e.g. "photon mass" or $ 2\rightarrow 2+\varepsilon $ regularization or else) and is well defined mathematically:
\begin{equation}
\mid T_{C+N}\mid_{q\neq0}^{2} = 4s^{2} S^{C} (q,q) + \int\frac{d^{2}q^{'}}{(2\pi)^{2}}\frac{d^{2}q^{''}}{(2\pi)^{2}} S^{C} (q^{'},q^{''})T_{N} (q-q^{'})T_{N}^{\ast} (q-q^{''})
\end{equation}
\[+4s \int\frac{d^{2}q^{'}}{(2\pi)^{2}} Im[S^{C} (q,q^{'})T_{N}^{\ast} (q-q^{'})]\]
where
\begin{equation}
S^{C} (q^{'},q^{''})= \int d^{2}b^{'}d^{2}b^{''} e^{i{q}^{'}{b}^{'}-i{q}^{''}{b}^{''}} e^{2i\alpha \Delta_{C} ( b^{'},\, b^{''})}
\end{equation}
and
\begin{equation}
\Delta_{C} ( b^{'},\, b^{''})= \frac{1}{2\pi}\int d^{2}k \frac{\mathcal{F}^{2}(k^{2})}{k^{2}}(e^{-ib^{''}k} - e^{-ib^{'}k})= 
\end{equation}
\[=\int_{0}^{\infty}\frac{dk}{k}\mathcal{F}^{2}(k^{2})[J_{0}(b^{''}k)-J_{0}(b^{'}k)].\]

In Eq.(4) we explicitly indicate the condition $ q\neq0 $ which corresponds to real experimental conditions (the scattered proton cannot be detected arbitrarily close to the beam axis). The "forward" observables , e.g. $ \sigma_{tot}(s) = Im T_{N}(s,0)/s $, are understood as the result of extrapolation $ t\rightarrow 0 $. However , this does not concern expressions appearing as integrands and able to contain terms like $ \delta (\textbf{q}) $.

In Eq.(6) the Coulomb singularity at $ k \rightarrow 0 $ is safely cured by the exponential (Bessel function) difference. 
Note that 
\begin{center}
$ S^{C} (q^{'},q^{''})\mid_{\alpha=0} = (2\pi)^{2} \delta (\textbf{q}^{'})(2\pi)^{2} \delta (\textbf{q}^{''})$
\end{center}
while
\begin{center}
$\int S^{C} (q^{'},q^{''}) d^{2}q^{'}d^{2}q^{''}/(2\pi)^{4}  = 1, \; \forall \alpha .$
\end{center}

In principle, when applying Eq.(4) to the data analysis, one could deal directly with Eq.(5) which is all-order (in $ \alpha $) exact expression free of singularities.

In unrealistic case of "electrically point like" nucleons, i.e. if $ \mathcal{F} = 1 $, we would have
a compact explicit expression for the Coulomb function $ S^{C} (q^{'},q^{''}) $ expressed in terms of the well known generalized functions described, e.g., in \cite{Vla}:
\begin{equation}
S^{C} (q^{'},q^{''}) =  (4\pi\alpha)^{2} \frac{(q^{''2}/q^{'2})^{i\alpha}}{q^{'2}q^{''2}}.
\end{equation}
However, it is hardly possible to obtain an explicit and "user friendly" expressions for arbitrary $ T_{N} $ and $ \mathcal{F} $.

Thus, in practice we have to use perturbative expansions in $ \alpha $. Let us notice, however,  that it would be a bit rash to limit to zero and first orders in $ \alpha $ because, e.g., the pure Coulomb contribution ($ \sim\alpha^{2} $) to the  observed $ d\sigma^{C+N}/dt $ at $ -t = \mathcal{O}(10^{-3} GeV^{2}) $ and $ \sqrt{s} = 13 TeV $ reaches near 30 \%. We notice that in relevant publications ( see e.g. Refs.\cite{Cahn} ) only the terms up to the first order in $ \alpha $ are retained in the amplitude, so when passing to the cross-section some terms are missing. This can lead to wrong estimation of parameters like $ \rho $ and so to wrong physical conclusions.

The basic kernel $ S^{C} (q^{'},q^{''}) $ has the following expansion in $ \alpha $  up to $ \alpha^{2} $ inclusively:
\begin{equation}
S^{C} (q^{'},q^{''})= (2\pi)^{2}\delta (\textbf{q}')(2\pi)^{2}\delta (\textbf{q}'')+2i\alpha (2\pi)^{3} [\hat{{\delta}_{C}}(q')\delta (\textbf{q}'')+ \hat{{\delta}_{C}}(q'')\delta (\textbf{q}')]+
\end{equation}
\[+ 2\alpha^{2}\pi^{2}\lbrace 2\hat{{\delta}_{C}}(q')\hat{{\delta}_{C}}(q'')-\delta (\textbf{q}')X(q'')- \delta (\textbf{q}'')X(q'))\rbrace + ...\]
where
\begin{equation}
\hat{{\delta}_{C}}(q)\doteq \int \frac{d\textbf{k}}{k^{2}}\mathcal{F}^{2}(k^{2})[\delta (\textbf{q})-\delta(\textbf{q}-\textbf{k})]
\end{equation}
and
\begin{equation}
X(q) = \int\frac{d\textbf{k}}{k^{2}}\mathcal{F}^{2}(k^{2})\int \frac{d\textbf{p}}{p^{2}}\mathcal{F}^{2}(p^{2})[\delta (\textbf{q}-\textbf{k}-\textbf{p}) -\delta (\textbf{q}-\textbf{k})-\delta (\textbf{q}-\textbf{p})+\delta(\textbf{q})].
\end{equation}
Quantities (9) and (10) are generalized functions which are defined on the space of appropriate test functions $ \phi (\textbf{q}) $. Normally infintely differentiable functions decreasing at infinity faster than any inverse power are used (Schwartz class $ S $) though in our case just differentiable and bounded at infinity functions would be fairly suitable.
Generalized functions (9) and (10) are defined as linear functionals  $ (...,\phi) $ with
\begin{equation}
(\hat{{\delta}_{C}},\phi) =\int \frac{d\textbf {k}}{k^{2}}\mathcal{F}^{2}(k^{2})( \phi(\textbf {k})-\phi(0)),
\end{equation}
and 
\begin{equation}
(X,\phi) = \int\frac{d\textbf{k}}{k^{2}}\mathcal{F}^{2}(k^{2})\int \frac{d\textbf{p}}{p^{2}}\mathcal{F}^{2}(p^{2})[\phi(\textbf{k}+\textbf{p}) -\phi(\textbf{k})-\phi (\textbf{p})+\phi(\textbf{0})].
\end{equation}
Distribution $ X $ can be expressed as a convolution  of the distribution $ \hat{{\delta}_{C}} $ with itself:
\begin{equation}
X(\textbf{q}) = (\hat{{\delta}_{C}}\star\hat{{\delta}_{C}})(\textbf{q})
\end{equation}
and in terms of local values we get
\begin{equation}
X(q)\mid _{q\neq 0} =  \frac{1}{q^{2}}\int \frac{dk^{2}dp^{2}}{k^{2}p^{2}} (-\lambda (q^{2},k^{2},p^{2}))_{+}^{-1/2}\times
\end{equation}
\[\times[q^{2}\mathcal{F}^{2}(k^{2})\mathcal{F}^{2}(p^{2})- (k^{2}\mathcal{F}^{2}(p^{2}) + p^{2}\mathcal{F}^{2}(k^{2}))\mathcal{F}^{2}(q^{2}) ]\]
where
\begin{center}
$ \lambda (q^{2},k^{2},p^{2})=q^{4}+k^{4}+p^{4} -2q^{2}k^{2} -2q^{2}p^{2} -2k^{2}p^{2} .$
\end{center}
and $ x_{+}^{\nu} \doteq x^{\nu},x\geq 0; = 0, x <0. $

One can readily see that the integrals in Eqs.(12)and (14) are well convergent at $ {k}^{2},{p}^{2} \rightarrow 0 $ . UV convergence is provided by the form factors as $ \mathcal{F}^{2}(k^{2})\sim k^{-8}$ at $ k^{2}\rightarrow \infty $.

Now we are able to write down the approximate
( up to $ \sim \alpha^{2} $ inclusively) expression(in units $ GeV^{-4}$) for the observed cross-section for pp scattering with account of Coulomb-nuclear interference ( $ t\approx -\textbf{q}^{2} $ and we do not explicitly indicate the $ s $-dependence in the amplitude):

\begin{equation}
16\pi s^{2} \frac{d\sigma_{C+N}^{pp}}{dt} = \mid T_{N}(q^{2})\mid^{2} + \alpha J_{1}+\alpha^{2} J_{2} + \mathcal{O}(\alpha^{3}).
\end{equation}
Here
\[J_{1} =  \lbrace \frac{16\pi s\mathcal{F}^{2}(q^{2})}{q^{2}} ReT_{N}({q}^{2})+\frac{2}{\pi}\int\frac{dk^{2}\mathcal{F}^{2}(k^{2})}{k^{2}} dq'^{2}(- \lambda (q^{2}, q'^{2}, k^{2}))^{-1/2}_{+} Im [T_{N}(q^{2})T^{\ast}_{N}(q'^{2})]\rbrace,\]
and then we break $ J_{2} $, in its turn, into three terms : 
\[J_{2} = J^{CC} _{2} + J^{CN}_{2} + J^{CNN}_{2},\]
where
$ J^{CC}_{2} $ is the term independent on the nuclear amplitude, $ J^{CN}_{2} $ the term linear in the nuclear amplitude, $ J^{CNN}_{2}$ the term quadratic in the  nuclear amplitude: 

\[J^{CC} _{2} = [\frac{8\pi s\mathcal{F}^{2}({q}^{2})}{q^{2}}]^{2},\]

 \[ J^{CN} _{2} = \frac{2sImT_{N}({q}^{2})}{q^{2}}\int \frac{dk^{2}dp^{2}}{k^{2}p^{2}}[q^{2}\mathcal{F}^{2}(k^{2})\mathcal{F}^{2}(p^{2})- (k^{2}\mathcal{F}^{2}(p^{2}) + p^{2}\mathcal{F}^{2}(k^{2}))\mathcal{F}^{2}(q^{2}) ]\times\]
  
  \[(-\lambda (q^{2},k^{2},p^{2}))_{+}^{-1/2}+\frac{4 s\mathcal{F}^{2}({q}^{2})}{q^{2}}\int\frac{dk^{2}dq'^{2}\mathcal{F}^{2}(k^{2})}{k^{2}} \times\]
\[\times(-\lambda (q^{2},k^{2},q'^{2}))_{+}^{-1/2}Im(T_{N} (q'^{2}) -T_{N} (q^{2})),\]

\[J^{CNN}_{2} = \: \mid\int \frac{d{k}^{2}\mathcal{F}^{2}((k^{2})}{2\pi k^{2}} dq'^{2}(-\lambda (q^{2},k^{2},p^{2}))_{+}^{-1/2}[T_{N} (q'^{2}) -T_{N} ({q^{2}})]   \mid^{2}\]
\[-\frac{1}{(2\pi)^{2}}\int\frac{d\textbf{k}\mathcal{F}^{2}(k^{2})}{k^{2}}\frac{d\textbf{p}\mathcal{F}^{2}(p^{2})}{p^{2}}[ReT_{N}(\textbf{q})( ReT_{N}(\textbf{q}-\textbf{p}-\textbf{k})\]
\[-ReT_{N}(\textbf{q}-\textbf{p})-ReT_{N}(\textbf{q}-\textbf{k})+ReT_{N}(\textbf{q}))+ ImT_{N}(\textbf{q})(ImT_{N}(\textbf{q}-\textbf{p}-\textbf{k})\]
\[-ImT_{N}(\textbf{q}-\textbf{p})-ImT_{N}(\textbf{q}-\textbf{k})+ImT_{N}(\textbf{q}))]\rbrace. \]

In order not to make Eq.(15) too 
unwieldy we have kept vector arguments in integration and in the scattering amplitudes in the last expression of the $ \alpha^{2} $ term.
To pass to invariant variables the integration measure $ d\textbf{k} d\textbf{p} $ is to be changed for 
$ dk^{2}dp^{2}dq'^{2}dq''^{2} (-\lambda (q^{2},q'^{2},k^{2} ))_{+}^{-1/2}(-\lambda (q^{2},q"^{2},p^{2} ))_{+}^{-1/2}$ and the following substitutions should be made:
\[T_{N}(\textbf{q})\rightarrow T_{N}(q^{2}),T_{N}(\textbf{q}-\textbf{p}-\textbf{k})\]
\[\rightarrow T_{N} (\frac{(q'^{2}-k^{2}-q^{2})(q''^{2}-p^{2}-q^{2})+(-\lambda (q^{2},q'^{2},k^{2} ))_{+}^{1/2}(-\lambda (q^{2},q''^{2},p^{2} ))_{+}^{1/2}}{2q^{2}} +\] \[+ \: q'^{2}+q''^{2}-q^{2}); \: T_{N}(\textbf{q}-\textbf{k})\rightarrow T_{N}(q'^{2}),\: T_{N}(\textbf{q}-\textbf{p})\rightarrow T_{N}(q''^{2}).\]

This expression is certainly quite bulky but we cannot avoid it if we keep $ \mathcal{O}(\alpha^{2}) $ terms which are important at low enough $ q^{2} $ characteristic for the region of CNI.
Plain fact is that it significantly differs from the expression that one obtains by taking the square of the CKL amplitude modulus (2),(3) used in Ref.\cite{TOT} for extraction of the $ \rho $ - parameter from the data. We believe that the application of our expression (15) given above can lead to essentially different values of $ \rho $ and, consequently, to different both numerical and conceptual conclusions.

 \section{ Conclusion and outlook}
 In this note we have exhibited a new, relatively simple but mathematically consisted, formula to deal with the Coulomb-nuclear interference which minimizes the use of IR regularizations and modifies the previously applied formula for $  T_{C+N} $.
 We also have shown that the usual retaining only the $ \mathcal{O}(\alpha) $ terms in the QED perturbative expansion of the amplitude $ T_{C+N} $ leads to loss of terms which can be important when passing to the cross-section and have explicitly calculated these terms. Their influence is potentially capable to change the values of the parameter $ \rho $ and, hence, the physical interpretation of the elastic proton-proton scattering at the LHC.
 Phenomenological application of the results presented here is the subject of a special publication \cite{Ezh} .
 \section{Acknowledgements}

 I am grateful to  Vladimir Ezhela, Anatolii Likhoded, Jan Ka\v{s}par, Vojtech Kundr\'{a}t, Per Grafstr\"{o}m, Roman Ryutin and Nikolai Tkachenko for their interest to this work and inspiring conversations and correspondence. I am  particularly indebted to Anatolii Samokhin for very fruitful discussions of some peculiar details of the paper as well as to the reviewer whose comments were helpful for improvement of  the presentation.

 This work is supported by the RFBR Grant 17-02-00120.

\end{document}